\documentclass[draft,grl]{agutex}

\usepackage[pdftex]{graphicx}

\setkeys{Gin}{draft=false}
\authorrunninghead{MURSULA ET AL.}

\titlerunninghead{SEASONAL SOLAR WIND SPEEDS}

\begin{document}

%
%

\title{Seasonal solar wind speeds for the last 100 years: 
Unique coronal hole structures during the peak and demise of the Grand Modern Maximum}

%

%
%


\authors{K. Mursula\altaffilmark{1}, L. Holappa\altaffilmark{1} and R. Lukianova\altaffilmark{2,3}} 
\altaffiltext{1}{ReSoLVE Centre of Excellence, Space Physics Research Unit, University of Oulu, Oulu, Finland}
\altaffiltext{2}{Geophysical Center of Russian Academy of Science, Moscow, Russia}
\altaffiltext{3}{Space Research Institute of RAS, Moscow, Russia}







%
%


\begin{abstract}

Solar coronal holes are sources of high-speed solar wind streams, which cause persistent geomagnetic activity especially at high latitudes.
Here we estimate seasonal solar wind speeds at 1 AU for the last 100 years using high-latitude geomagnetic measurements and show that they give information on the long-term evolution of important structures of the solar large-scale magnetic field,  
such as persistent coronal holes.
We find that the centennial evolution of solar wind speed at 1 AU is different for equinoxes and solstices, reflecting differences in the evolution of polar coronal hole extensions and isolated low-latitude coronal holes.
Equinoctial solar wind speeds had their centennial maximum in 1952, during the declining phase of solar cycle 18, verifying that polar coronal holes had exceptionally persistent extensions just before the peak of the Grand Modern Maximum of solar activity.
On the other hand, solstice speeds had their centennial maximum during the declining phase of solar cycle 23 due to large low-latitude coronal holes.
A similar configuration of seasonal speeds as in cycle 23 was not found earlier, not even during the less active cycles of early 20th century.
Therefore the exceptional occurrence of persistent, isolated low-latitude coronal holes in cycle 23 is not related to the absolute level of sunspot activity but, most likely, to the demise of the Grand Modern Maximum.



\end{abstract}

%
%

%

\begin{article}

%
%

\section{Introduction}

The large-scale evolution of solar wind (SW) and the interplanetary magnetic field is determined by the solar magnetic field, whose structure varies over the solar cycle \citep{Wang_1991, McComas_2008, Pinto_2011}. 
The highest values of solar wind speed ($V$) at the Earth's orbit are observed during the declining phase of the solar cycle, when high-speed streams (HSS) from the equatorward extensions of polar coronal holes extend to low heliographic latitudes \citep{Krieger_1973, Gosling_1976}. 
During their passage through the interplanetary space, HSS compress the ambient slow solar wind plasma and the interplanetary magnetic field, which may lead to the formation of corotating interaction regions (CIR).  
Coronal mass ejections and CIR/HSS hitting the Earth's magnetic field are the main drivers of geomagnetic activity.
While coronal mass ejections produce the majority of strong geomagnetic storms, CIR/HSS produce a large fraction of storms of minor to moderate intensity \citep{Richardson_2006, Zhang_2007}.
However, the combination of high solar wind speed and Alfv\'en waves embedded in HSS produce strong and persistent auroral activity \citep{Tsurutani_2006}. 
The annual occurrence of substorms is also strongly modulated by the occurrence of HSS \citep{Tanskanen_2005, Tanskanen_2011}. 

High-speed streams have the largest relative effect on geomagnetic activity at high latitudes \citep{Finch_2008, Lukianova_2012, Holappa_2014}, and high-latitude geomagnetic activity has been exploited to find information about the historical occurrence of HSS. 
\citet{Mursula_2015} used observations at two high-latitude magnetic stations (Godhavn/Qeqertarsuaq, and Sodankyl\"a) to reconstruct the annual means of solar wind speed proxies over the last 100 years, covering the special time interval of the Grand Modern Maximum of solar activity \citep{Solanki_2000}. 
They found that a short period of typically 1-2 years of high values of annual SW speed occurred in the declining phase of each of the eight studied solar cycles (16--23). 
During the declining phase of cycle 18 in the early 1950s, high HSS activity continued exceptionally long, for three successive years, with the highest annual speed found in 1952. 
\citet{Mursula_2015} noted that cycle 19, which marks the sunspot maximum of the Grand Modern Maximum, was most likely preceded by exceptionally strong polar fields produced by frequent surges of new flux leading to persistent extensions of polar coronal holes during the declining phase of the previous cycle.
This relation verifies one of the two basic tenets of solar dynamo theory of the poloidal-to-toroidal field transition (so called $\Omega$ mechanism) \citep{Babcock_1961} for the period of the highest known solar activity. 

While the annual averages of solar wind speed give important insight to the long-term evolution of solar magnetic fields and coronal holes, a study using higher time resolution is useful because the typical lifetime of coronal holes is several solar rotations but less than a year [\citet{Harvey_1979}; \citeauthor{Mursula_Zieger_1996}, \citeyear{Mursula_Zieger_1996, Mursula_2001}]. 
Because of the $7.2^{\circ}$ annual variation of the Earth's heliographic latitude, HSS from polar coronal holes (or their extensions) reach the ecliptic plane more often close to equinoxes.
On the other hand, isolated equatorial or low-latitude coronal holes would lead to a larger relative fraction of Earth-bound HSS close to solstices.
Thus, the differences in the long-term evolution of HSS between the seasons corresponding to low vs. high heliographic latitudes can be used to study the occurrence and the approximate location of persistent coronal holes. 
In this paper we study the long-term evolution of HSS at seasonal time resolution during the last 100 years (1914-2014). 
The paper is organized as follows. 
In Section 2 we introduce the solar wind speed proxy and study its relation to the measured solar wind speed. 
In Section 3 we derive the seasonal solar wind speeds in 1914-2014. 
Finally, we discuss the implications of our results for the evolution of solar magnetic fields and coronal holes and give our conclusions in Section 4.

\section{Extracting SW speeds from high-latitude disturbances}

In order to obtain a long-term proxy of the seasonal SW speed we use the horizontal magnetic field measurements at Sodankyl\"a observatory (geographic: $67.37^{\circ}$ latitude, $26.63^{\circ}$ longitude; geomagnetic: $64^{\circ}$ lat and $119^{\circ}$ long), located near the equatorward boundary of the auroral oval. At Sodankyl\"a, recordings of the Earth's magnetic field vector have been made since 1914 (interrupted only by the World War II in 1945), forming the longest series of high-latitude geomagnetic measurements.

The largest perturbations at Sodankyl\"a occur in the horizontal ($H$) geomagnetic component due to the auroral electrojets. 
During the growth and expansion phases of magnetospheric substorms, the westward auroral electrojet intensifies and expands, and decays back to the quiet-time level during the recovery phase \citep{Akasofu_1964}. 
The strongest westward currents during substorms are due to the substorm current wedge located around the midnight sector \citep{Clauer_McPherron_1974}.  
It has been shown that the substorm-related disturbances in the midnight sector are indeed dominated by HSS \citep{Tanskanen_2005}, while HSS have a smaller relative effect at other local times \citep{Finch_2008}.
Therefore, in order to have the best possible proxy for the solar wind speed we use the data from the 20-23 UT (all four hours included) time interval, corresponding to 22-01 local time at Sodankyl\"a.

We use the geomagnetic disturbance parameter $\Delta{H}$ as a measure of westward auroral electrojet intensity.
We define $\Delta{H}$ as follows. 
For each month we calculate the quiet-time level $H(q)$ as the average of four-hourly night-time (22-01 local time) $H$ values during the five quietest days of the month. 
(The quietest days have been determined from the local $K$ indices and are available at Sodankyl\"a observatory web page). 
Then we calculate the monthly disturbance $\Delta{H}$ by subtracting the monthly average of night-time $H$ values of all days from $H(q)$.
(Since disturbances reduce the H-component, $\Delta{H}$ will always be positive.) 
Then we calculate the seasonal (three-month) averages of the monthly $\Delta{H}$ values for spring (Feb-Apr), summer (May-Jul), fall (Aug-Oct) and winter (Nov-Jan).
(Here seasons are mainly ordered by the Earth's heliographic latitude).

We also use the hourly means of solar wind speed $V$ in 1964-2014 from the OMNI data base (http://omniweb.gsfc.nasa.gov/).
While there are practically no data gaps in the Sodankyl\"a $H$ values during the space age (only 11 hourly means missing since 1964), there are numerous data gaps in the OMNI data base especially during 1980s and early 1990s. 
Therefore, we calculate seasonal means using only those night hours when simultaneous measurements of both $\Delta{H}$ and $V$ exist. 
In order to have sufficient statistics, we neglect those seasonal values where the data coverage remains smaller than 30\%.

\begin{figure}
\includegraphics[width=0.90\linewidth]{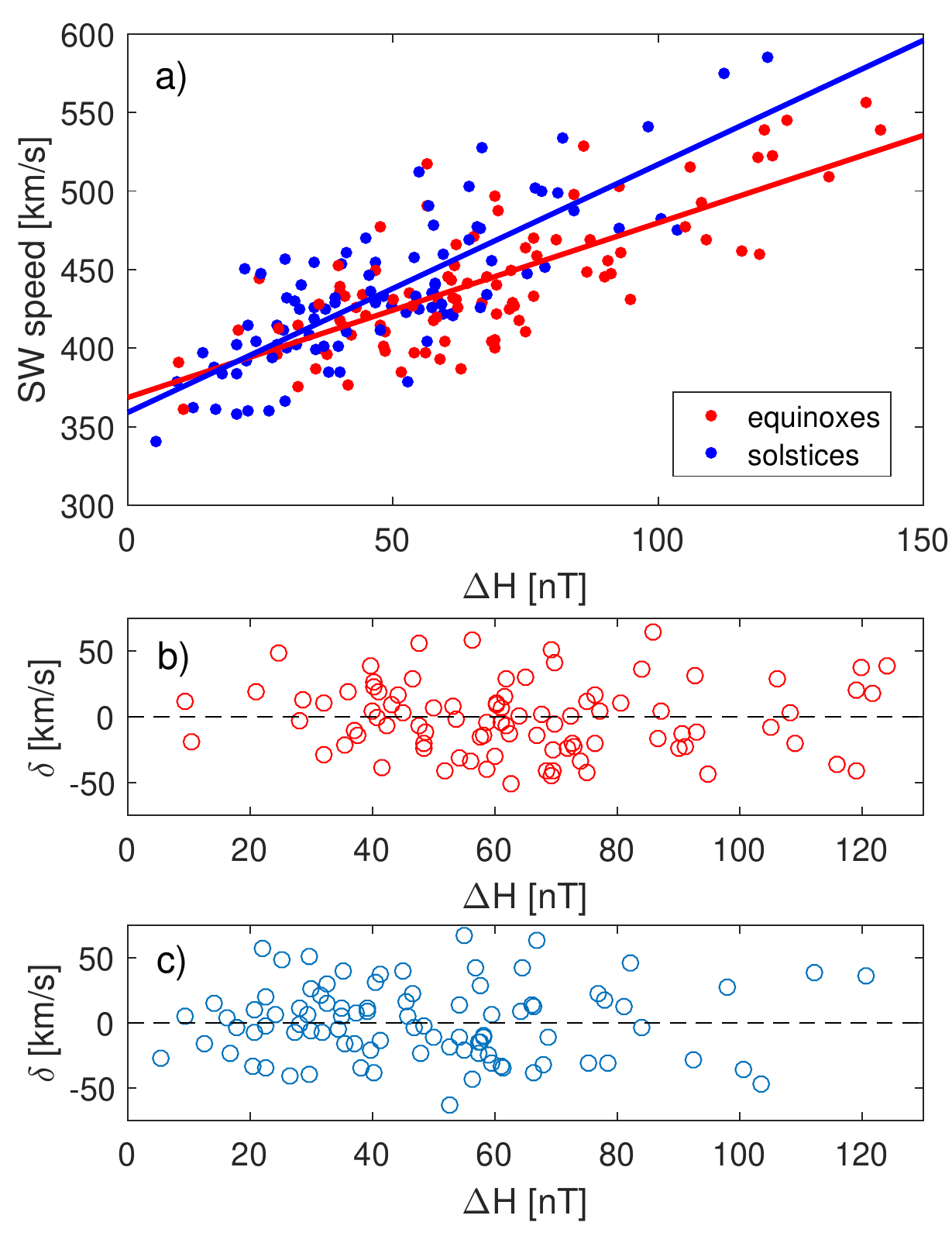}
\caption{a) Scatter plot between seasonal means of $\Delta{H}$ and solar wind speed during equinoxes (red dots) and solstices (blue bots), together with their best-fit lines. b) Residuals of regressions as a function of $\Delta{H}$ for equinoxes and c) solstices.}
\label{dh_sw_corr}
\end{figure}

Figure \ref{dh_sw_corr}a shows a scatter plot between the solstice (summer and winter seasons) and equinox (spring and fall) means of $\Delta{H}$ and the measured seasonal solar wind speed $V$ in 1964-2014, together with the best-fit regression lines. 
Table 1 gives the regression parameters (slope, intercept), correlation coefficients and the corresponding p-values (using first-order autoregressive AR(1) noise model) for the two fits.
One can see that $V$ increases more steeply as a function of $\Delta{H}$ during solstices than during equinoxes.
Thus, a fixed value of $V$ produces a larger disturbance $\Delta{H}$ during equinoxes than solstices. 
On the other hand, intercepts are closely similar, indicating a similar quiet-time level.
This difference is related to the well known semiannual variation of geomagnetic activity, which is mainly due to the equinoctial \citep{Cliver_2000, Lyatsky_2001} and Russell-McPherron \citep{Russell_1973} effects modulating the strength of solar wind-magnetosphere coupling.
Figures \ref{dh_sw_corr}b and \ref{dh_sw_corr}c show the fit residuals $\delta$ = $V$(measured) - $V$(predicted) as a function of $\Delta{H}$ for equinoxes and solstices, respectively.
Correlations are highly significant in both seasons (slightly better in solstices) and no obvious outliers are seen.
Moreover, the residuals lie symmetrically around zero for the whole range of $\Delta{H}$, indicating a homoscedastic distribution of errors.  
Thus, the standard least squares fit performs well and can be reliably used to reconstruct  seasonal proxies for the solar wind speed. 


\section{Seasonal SW speeds in 1914-2014}

\begin{figure}
\includegraphics[width=0.9\linewidth]{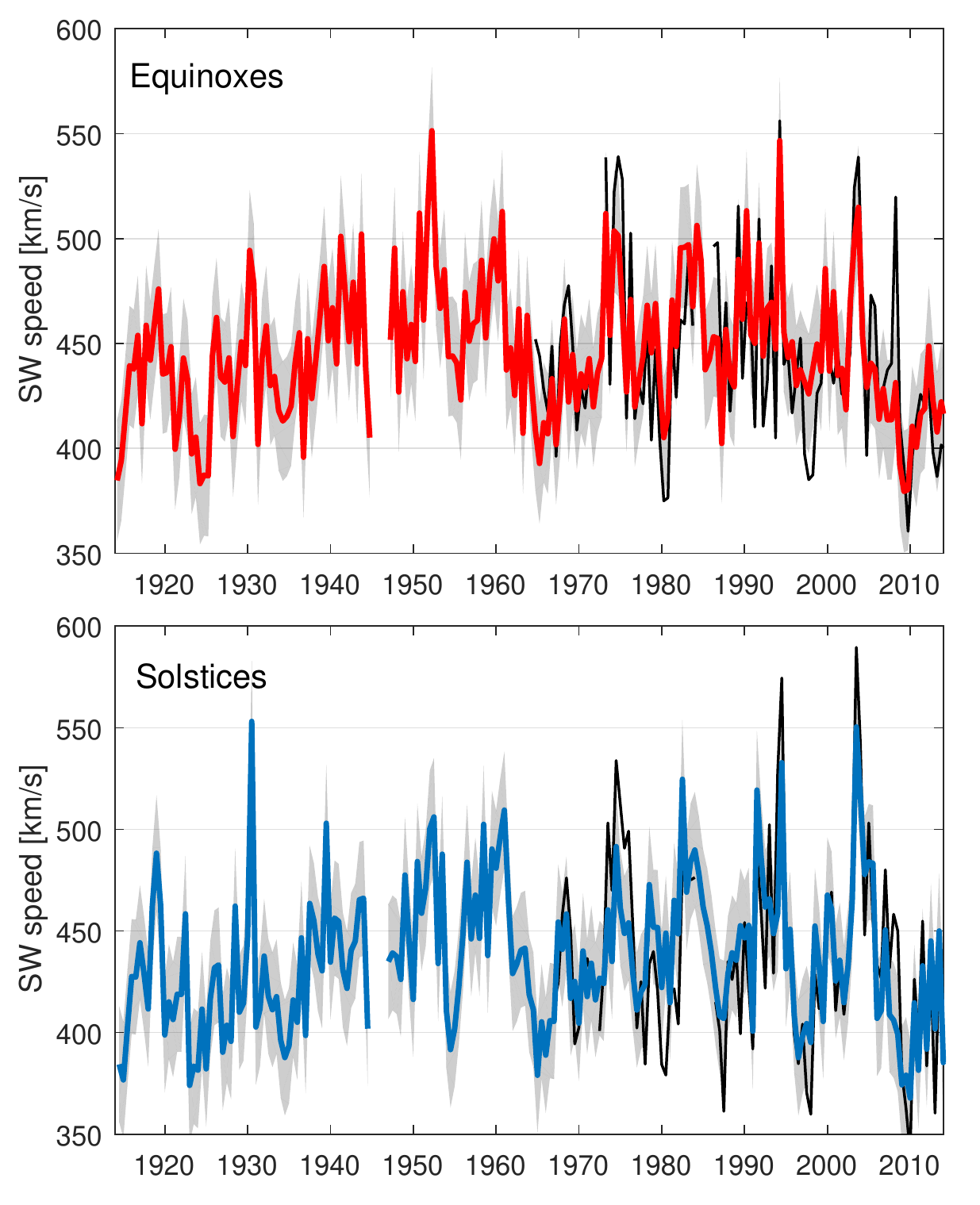}
\caption{Solar wind speed proxies for a) equinoxes (red line) and b) solstices (blue line) with $\pm 1\sigma$ error estimates (shaded gray areas). Black lines denote the measured seasonal means of solar wind speed.}
\label{seasonal_V_recons}
\end{figure}

The above discussed seasonal regressions between $\Delta{H}$ and the measured SW speed in 1964-2014 can now be applied to  reconstruct the seasonal SW speeds for the pre-satellite era since 1914. 
Figures \ref{seasonal_V_recons}a and \ref{seasonal_V_recons}b show the SW speed proxies in 1914-2014 for equinoxes and solstices, respectively, together with the measured seasonal solar wind speeds in 1964-2014. 
Here all available data (not only simultaneous data with the solar wind measurements) are used when calculating measured and estimated SW speeds.
Accordingly, Figure \ref{seasonal_V_recons} presents the SW speed proxies for the case of gapless measurement of SW speeds. 
This slightly increases the differences between the measured and estimated SW speeds during the common time interval 1964-2014. 
However, this is unavoidable and irrelevant for this paper where the main purpose is to compare the two seasonal SW speed proxies.

Figure \ref{seasonal_V_recons} shows that there are some significant and interesting differences between the SW speed proxies for equinoxes and solstices.
Low ($<400$ km/s) seasonal means of solar wind speed occur more often during solstices than during equinoxes, which can be seen both in the measured and proxy values especially in recent decades (since 1990s). 
The highest equinox maxima occurred in 1952 and 1994 while the highest solstice maxima occurred in 1930 and 2003. 
Because of these differently timed maxima, the long-term evolution of solar wind speed is quite different during equinoxes and solstices.
(Note also that SW speed was not particularly high around 1960, in the declining phase of the all-time high sunspot cycle 19, ahead of the low sunspot cycle 20).

\begin{figure}
\includegraphics[width=\linewidth]{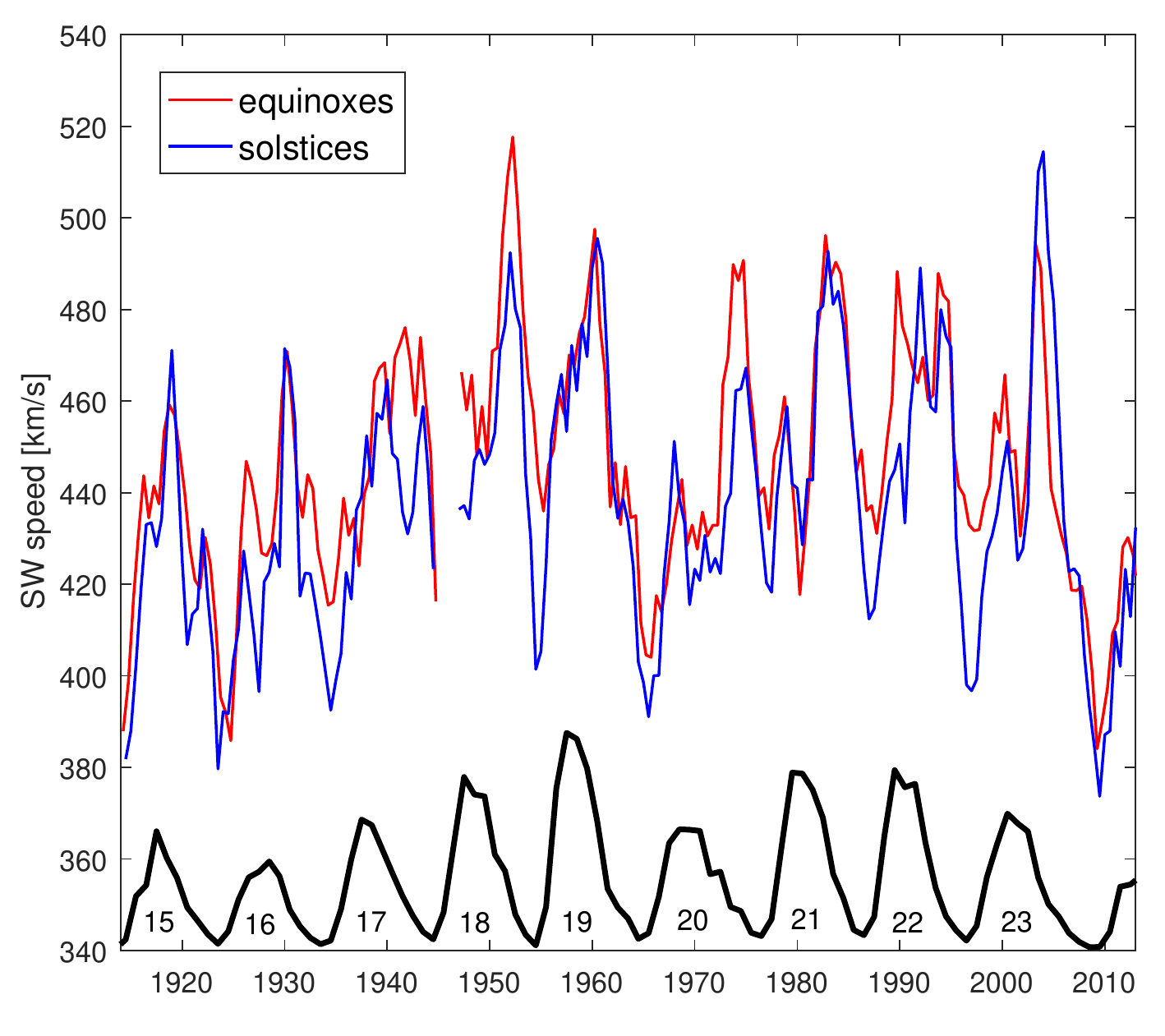}
\caption{Three-point running means of the reconstructed solar winds speeds for equinoxes and solstices. Annual sunspot number (arbitrary scale) and solar cycle numbers are included for reference.}
\label{filt}
\end{figure}

The peaks in Figure \ref{seasonal_V_recons} correspond to individual seasonal (three-monthly) means, thus reflecting a rather short-term occurrence of coronal holes. 
We also want to study the occurrence of the most persistent coronal hole structures lasting for longer than one season.
For that, we show in Figure \ref{filt} the three-point running means of the seasonal SW speeds of Figure \ref{seasonal_V_recons}, separately for equinox and solstice proxies. 
Thus, each point in Figure \ref{filt} gives an average over 1.5 years.
Figure \ref{filt} shows that some of the peaks in Figure \ref{seasonal_V_recons}, in particular the solstice peak in 1930 was based only on one high (summer) solstice value, and lost in relative height in Figure \ref{filt}.
Others, in particular the equinox peak in 1952 and the solstice peak in 2003 correspond to the most persistent coronal hole activity and retain their all-time (centennial) high peak status during the respective season in Figure \ref{filt}.

Figure \ref{filt} shows interesting similarities and differences in the long-term evolution of the reconstructed SW speed between equinoxes and solstices and, thereby, between polar coronal hole extensions and low-latitude coronal holes. 
During the low cycles 15 and 16, SW speeds attained roughly similar peak values of about 460-470 km/s in both seasons. 
During the next, more active cycle 17, the peak values remained still at a fairly similar level but both the solstice speed and especially the equinox speed depict 2-3 peaks of roughly equal height.
The two curves are then separated in cycle 18, when the equinox speeds attain their all-time maximum value of about 520 km/s. 
Note that the solstice peak in cycle 18 and in most remaining cycles (19, 21, 22), as well as all equinox peaks since cycle 19 reach a roughly similar level of about 490-500 km/s, which is somewhat higher than the level of peaks during the low cycles 15-17.
During cycle 20 the solstice peak remains at the low level of the early cycles.
Interestingly, both speeds depict multiple peaks during cycle 22, just before the solstice speed reaches its all-time maximum in cycle 23. 
This evolution resembles the evolution of speeds during cycles 17-18 before the all-time peak in equinox speed.
Note also that the two all-time maxima attain surprisingly similar values.

Figure \ref{filt} also shows that during solar minima the solar wind speed is systematically lower during solstices than equinoxes. 
This is expected because the Earth is within the slow solar wind of the streamer belt more often during solstices than during equinoxes.
(This was seen also for the unsmoothed seasonal solar wind speeds in Figure \ref{seasonal_V_recons}.)
Note also that while the equinox and solstice minima are quite similar during the minima between cycles 14-15 and 15-16, the difference is significantly larger during the minima of the more active cycles.
The level of equinox minima is considerably raised during the more active cycles, but also the solstice minima attains higher values.
This indicates that the streamer belt was rather stable and wide during the minima of the low cycles, but thinner and probably also more tilted during the minima of more active cycles.
(Note that larger activity tends to increase longitudinal asymmetry and, thereby, lead to a larger tilt).

The recent very deep minimum in 2009 can be seen both in solstice and equinox speeds.
During this minimum all equatorial and low-latitude coronal holes (and polar coronal hole extensions) disappeared \citep{Tsurutani_2011}, leading to the slowest solar wind speed during the space age. 
Interestingly, Figure \ref{filt} shows that the level of persistent solar wind speeds during this minimum is almost exactly the same as during the first two minima, indicating a return of the coronal and solar wind speed conditions to a similar situation as 100 years ago.

\section{Discussion and Conclusions}

In this paper we have used the longest available set of high-latitude geomagnetic observations to reconstruct a local measure ($\Delta{H}$) of geomagnetic disturbance level which is suitable to study the long-term evolution of Earth-bound solar wind speed.
We have quantified the relation between $\Delta{H}$ and the measured solar wind speed separately during equinoxes and solstices, finding high correlations for both seasons. 
We have estimated the seasonal means of the solar wind speed in 1914-2014 separately for equinoxes and solstices.
Even though the $7.2^{\circ}$ inclination of the Earth's orbit with respect to the heliographic equator is rather small, it is large enough to depict long-term variations in the latitudinal distribution of coronal holes, especially, in the relative effective fraction of polar coronal hole extensions and isolated low-latitude coronal holes.
 
Interestingly, we find that the long-term evolution of solar wind speed is quite different during equinoxes and solstices.
While the solar wind speed during equinoxes shows a centennial maximum in 1952, during the declining phase of cycle 18, the solstice speeds maximise in 2003, during the declining phase of cycle 23.
These two years were also the highest peaks in annual solar wind speed proxies during the last 100 years \citep{Mursula_2015}.
We have found here that, with increasing sunspot activity since cycle 17, extensions of polar coronal holes (and thereby polar fields) increase rapidly, leading to the centennial maximum of equinoctial solar wind speed during cycle 18.
We suggested earlier that the exceptionally high solar wind speed in 1952 reflects the persistent existence of equatorward extensions of polar coronal holes, signaling the build-up of strong polar magnetic fields during the declining phase of cycle 18 \citep{Mursula_2015}.
However, this explanation does not apply to the peak in 2003, since the polar fields remained rather weak during the declining phase of solar cycle 23 and the minimum thereafter \citep{Smith_2008, Wang_2009}.

We have shown here that the solar wind speed peak in 2003 was the centennial maximum during solstices (but not during equinoxes), indicating that the HSS observed at 1 AU in 2003 largely emanated from isolated low-latitude coronal holes rather than from polar coronal hole extensions.
It is known that exceptionally persistent, isolated low-latitude coronal holes existed during the declining phase of cycle 23 \citep{Gibson_2009, Fujiki_2016}. 
So, cycle 23 was unique in persistent low-latitude coronal hole activity during the last 100 years.
Interestingly, no equally persistent low-latitude coronal holes are found during last 100 years, including the very low-active cycles in the beginning of the 20th century. 
One might expect that during such low cycles, when polar fields remain weak and polar coronal holes rather limited, low-latitude coronal holes and related Earth-bound high-speed streams could be more typical than during the more active cycles.
However, this is not the case. 
Thus, the coronal hole evolution in cycle 23 was unique during the last 100 years, probably reflecting the demise of the Grand Modern Maximum.
It is likely that this change in coronal holes is related to the other unique features of cycle 23, such as the  exceptionally weak polar magnetic fields \citep{Smith_2008} and the changes in the distribution of magnetic fields \citep{Penn_2006} and sizes \citep{Lefevre_2011, Clette_2012} of sunspots, and in their relation to several other solar parameters \citep{Lukianova_2011}.
Only later work will reveal if and how all these observed, unique features of cycle 23 are related to each other and to the corresponding changes in global solar magnetic fields.

While the solar wind speed minima of low-activity cycles attain quite similar, very low values in equinoxes and solstices, the minimum speeds of more active cycles are higher in both seasons. 
Moreover, the equinoctial speed minima increase faster than solstice speed minima when the overall solar activity increases.
These results give strong evidence that the properties of the streamer belt also varied with long-term solar activity during the last 100 years.


%

%
%
%
%
%

%
%
%
%

\begin{acknowledgments}
We acknowledge the financial support by the Academy of Finland to the ReSoLVE Centre of Excellence (project no. 272157).
We thank the Sodankyl\"a Geophysical Observatory for providing the magnetic field data at (\texttt{http://www.sgo.fi/}).
The solar wind data were downloaded from the OMNI2 database (\texttt{http://omniweb.gsfc.nasa.gov/}).
\end{acknowledgments}

\end{article}



%
%
%
%
%
%

\newpage

\begin{table}
\begin{center}
\begin{tabular}{l|cc}
   &  Equinoxes  & Solstices \\\hline
slope $a$ [km/s/nT] & 1.11 & 1.58  \\\hline
intercept $b$ [km/s] & 369 & 359  \\\hline
cc & 0.74 & 0.80 \\\hline
p-value & $< 10^{-11}$ & $< 10^{-8}$ \\\hline
\end{tabular}\caption{Regression parameters (slope $a$ intercept $b$), correlation coefficients (cc) and p-values for equinoxes and solstices.}
\label{table}
\end{center}
\end{table}


\newpage

\clearpage

\end{document}